\documentclass[11pt]{article}
\usepackage{graphicx}
\usepackage{hyperref}
\usepackage{amssymb}
\usepackage{authblk}

\usepackage[usenames, dvipsnames]{color}

\usepackage{amsmath}

\newcommand{\unitx}{\,\mathrm}

\date{\today}

\title{Search for Dark Matter Annihilation in the Earth using the ANTARES Neutrino Telescope}

\author[1]{A.~Albert}
\author[2]{M.~Andr\'e}
\author[3]{M.~Anghinolfi}
\author[4]{G.~Anton}
\author[5]{M.~Ardid}
\author[6]{J.-J.~Aubert}
\author[7]{T.~Avgitas}
\author[7]{B.~Baret}
\author[8]{J.~Barrios-Mart\'{\i}}
\author[9]{S.~Basa}
% ...
\affil[1]{\scriptsize{GRPHE - Universit\'e de Haute Alsace - Institut universitaire de technologie de Colmar, 34 rue du Grillenbreit BP 50568 - 68008 Colmar, France}}
\affil[2]{\scriptsize{Technical University of Catalonia, Laboratory of Applied Bioacoustics, Rambla Exposici\'o, 08800 Vilanova i la Geltr\'u,Barcelona, Spain}}
\affil[3]{\scriptsize{INFN - Sezione di Genova, Via Dodecaneso 33, 16146 Genova, Italy}}
\affil[4]{\scriptsize{Friedrich-Alexander-Universit\"at Erlangen-N\"urnberg, Erlangen Centre for Astroparticle Physics, Erwin-Rommel-Str. 1, 91058 Erlangen, Germany}}
\affil[5]{\scriptsize{Institut d'Investigaci\'o per a la Gesti\'o Integrada de les Zones Costaneres (IGIC) - Universitat Polit\`ecnica de Val\`encia. C/  Paranimf 1, 46730 Gandia, Spain.}}
\affil[6]{\scriptsize{Aix-Marseille Universit\'e, CNRS/IN2P3, CPPM UMR 7346, 13288 Marseille, France}}
\affil[7]{\scriptsize{APC, Universit\'e Paris Diderot, CNRS/IN2P3, CEA/IRFU, Observatoire de Paris, Sorbonne Paris Cit\'e, 75205 Paris, France}}
\affil[8]{\scriptsize{IFIC - Instituto de F\'isica Corpuscular (CSIC - Universitat de Val\`encia) c/ Catedr\'atico Jos\'e Beltr\'an, 2 E-46980 Paterna, Valencia, Spain}}
\affil[9]{\scriptsize{LAM - Laboratoire d'Astrophysique de Marseille, P\^ole de l'\'Etoile Site de Ch\^ateau-Gombert, rue Fr\'ed\'eric Joliot-Curie 38,  13388 Marseille Cedex 13, France}}
% ...

\begin{document}

\maketitle

\begin{abstract}
A search for a neutrino signal from WIMP pair annihilations in the centre of the Earth has been performed with the data collected with the ANTARES neutrino telescope from 2007 to 2012. The event selection criteria have been developed and tuned to maximise the sensitivity of the experiment to such a neutrino signal. No significant excess of neutrinos over the expected background has been observed. Upper limits at $90\%$ C.L. on the WIMP annihilation rate in the Earth and the spin independent scattering cross-section of WIMPs to nucleons $\sigma^{SI}_p$ were calculated for WIMP pair annihilations into either $\tau^{+}\tau^{-}$,  $W^+W^-$, $b\overline{b}$ or the non-SUSY $\nu_{\mu}\bar{\nu}_{\mu}$ as a function of the WIMP mass (between $25\unitx{GeV/c^2}$ and $1000\unitx{GeV/c^2}$) and as a function of the thermally averaged annihilation cross section times velocity $\langle\sigma_{A} v\rangle_{Earth}$ of the WIMPs in the centre of the Earth. For masses of the WIMP close to the mass of iron nuclei ($50\unitx{GeV/c^2}$), the obtained limits on $\sigma^{SI}_p$ are more stringent than those obtained by other indirect searches.
\end{abstract}

\section{Introduction}\label{ch_1}

The Universe consists of a large fraction of dark matter (DM)\cite{bullet}\cite{dm_rot_curves}\cite{DM_sm1}.
DM particles do not interact electromagnetically, are stable on cosmological time scales, cannot be dominantly baryonic, and must move with non-relativistic speeds already at the structure formation epoch. 
The DM relic abundance today as a result of thermal production requires a particle with a thermally averaged annihilation cross-section of about
\begin{equation}\label{eq:sa}
\langle\sigma_A v\rangle = 3 \cdot 10^{-26}\quad \mathrm{cm}^3 \mathrm{s}^{-1} \ ,
\end{equation}
which is the natural scale at which a weakly-interacting particle \cite{SDM} would be expected.
The hypothetical Weakly Interacting Massive Particles (WIMPs) are therefore widely regarded as excellent DM candidates. Such particles arise in different theories, such as supersymmetric (SUSY) models \cite{SDM} (the fact that SUSY predicts a particle with the right properties is often referred to as the `WIMP miracle') or models with extra dimensions \cite{Kaluza2}. 
WIMPs from supersymmetric models, such as the Minimal Supersymmetric Extension of the Standard Model are widely regarded as the most promising dark matter candidates. In most cases the lightest supersymmetric particle is the lightest neutralino. 

WIMPs can be detected either in collider experiments by observing missing energy and momentum in particle collisions, directly via the observation of the nuclear recoils from the scattering of WIMPs off nuclei \cite{XENON100-2}\cite{lux_results}\cite{Edelweiss}\cite{PandaX} or indirectly \cite{DM_sm2} via the observation of products from WIMP self-annihilations.

Most indirect experiments rely on the fact that DM particles present in the Galactic halo may lose energy by interacting with nuclei of massive objects, as for example the Sun and the Earth itself, and may accumulate in the centre of these bodies under their gravitational potential. As shown in Section \ref{ch_2}, the accumulated DM particles may then self-annihilate. Among the final-state particles of the decay products, neutrinos can almost freely escape the massive objects, reaching neutrino telescopes located near the surface of the Earth.
The energy spectrum of the produced neutrino flux depends on the specific nature of DM particles \cite{DM_sm3} (in the following, the WIMP scenario will be assumed), the DM annihilation channel and mass. The expected neutrino event rates are also a function of the DM local density and velocity distribution and of the chemical composition of the celestial trapping object.

Searches for neutrinos from the direction of the Sun \cite{antares_dmsun}\cite{ICsun1}\cite{ICsun2}\cite{secluded}\cite{Baksan} of the Galactic Centre \cite{ANTARES_GC}\cite{ICgc}
and of the Earth core \cite{icecube_theory}\cite{icecubeearth} have already been carried out by neutrino telescopes and other neutrino experiments \cite{kamiokande}\cite{macro}.

WIMPs become gravitationally bound to the Earth if their velocity is smaller than the escape velocity from Earth, which ranges from $ 11.1\unitx{km/s}$ to $14.8\unitx{km/s}$ (at the surface and at the centre respectively). 
The velocity of WIMPs follows a Maxwell-Boltzmann distribution, the canonical value for the velocity dispersion is $270\unitx{km/s}$ (this value is subject to considerable uncertainty \cite{SDM}).
Under these conditions, only a small fraction of WIMPs would lose enough energy to become captured if there is a large difference between the mass of the WIMP and the mass of the nucleus the particle is scattering on. 
Capture of WIMPs in the Earth is expected to be dominated by spin-independent elastic scattering on the most abundant heavy nuclei, mainly iron and nickel.

In this paper, an indirect search for DM towards the centre of the Earth using data collected in 2007 -- 2012 by the ANTARES neutrino telescope is presented.
In section \ref{ch_2}, the WIMP capture process in the Earth is explained and quantified. In section \ref{ch_3}, 
the ANTARES neutrino telescope, the background events and the potential signal events for this search are presented. 
The event reconstruction methods, the selection criteria and their optimisation are described in section \ref{ch_4}.
In section \ref{ch_5}, the results of the analysis are presented and discussed.

\section{Capture and Annihilation of WIMPs in the Earth}\label{ch_2}
The process of WIMP annihilation in the centre of the Earth produces standard model particles (such as $W^+W^-$, $\tau^+\tau^-$, $b\overline b$ pairs) that include neutrinos in their final-state decay products. Muon neutrinos (in the following, `neutrinos' refers to the sum $\nu_\mu+\overline \nu_\mu$) can be detected via up-going muons from their interaction with matter.

According to \cite{SDM}, the WIMP annihilation rate $\Gamma_A(t)$ in the Earth can be written as (here and in the following, $c$ was set to 1):
\begin{equation}
\Gamma_A(t) = \frac{1}{2} C_A  N^2(t) = \frac{1}{2} C_C \tanh^2\left(\frac{t}{\tau} \right) \ , \quad \tau=\frac{1}{\sqrt{C_C C_A}}  .
\label{eq:A_r}	
\end{equation}
Here $N(t)$ is the total number of WIMPs at time $t$ after the formation of the Earth. 
The equilibrium time scale $\tau$ determines the time needed for WIMPs to reach equilibrium between capture and annihilation in the core of an astrophysical object. It depends on the annihilation factor $C_A$ and on the capture factor $C_C$. It can be shown that equilibrium is generally not reached in the case of Earth. $C_A$ is defined \cite{SRN} as
\begin{equation}
C_A=\frac{\langle\sigma_A v\rangle_{Earth}}{V_0} \left( \frac{m_{\chi}}{20\unitx{GeV}} \right) ^\frac{3}{2} .
\label{eq:C_A}
\end{equation}
Here $\langle\sigma_A v\rangle_{Earth}$ is the thermally averaged annihilation cross-section times speed, $m_{\chi}$ is the WIMP mass and ${V_0}$ is the effective volume of the Earth, taken from \cite{SRN}. 
The capture factor $C_{C}$ depends on the unknown WIMP mass and cross-section for interactions with Earth nuclei, the velocity of WIMPs in the halo and their local mass density. It can be written as
\begin{equation}
C_{C}=\frac{\sigma^{SI}_p \rho_{0.3}^{\chi}}{ m_{\chi} \bar{v}_{270}} \sum\limits_{i} F_i^*(m_{\chi}) \ .
\label{eq:C_C}
\end{equation}
The local halo mass density $\rho_{0.3}^{\chi}$ is estimated from observations and it is expressed in units of $0.3\unitx{GeV/cm^3}$;
the WIMP velocity dispersion $\bar{v}_{270}$, expressed in units of $270\unitx{km/s}$, can be estimated through simulations.
The WIMP cross-section depends on the chemical composition of the Earth and on the scattering cross-section of WIMPs to protons and neutrons.
The dominant process is due to spin-independent elastic scattering of WIMPs to nucleons, whose cross-section is usually referred to that of the WIMPs to protons, denoted as $\sigma^{SI}_{p}$. 
In fact, for neutralinos and most other WIMP candidates, the spin-independent scattering cross-sections on protons and neutrons are roughly identical \cite{proton-neutron}. 
The Earth composition enters in the factors $ F_i^*(m_{\chi})$ and whose sum is taken over all kinds of nuclei present in the Earth. 
 
\begin{figure}[tbh]
\centering
\includegraphics[angle=0,width=12cm]{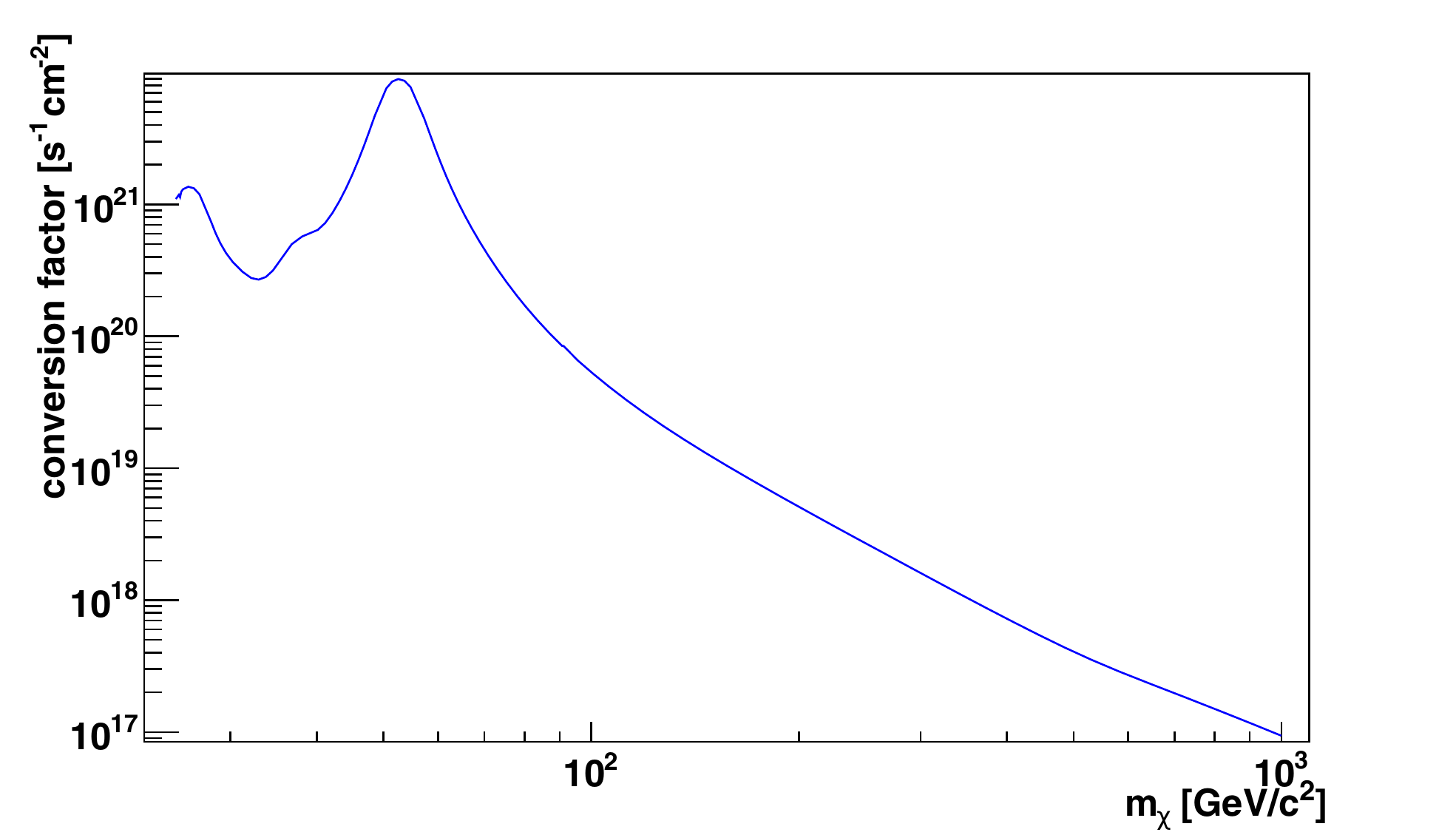}
\caption{The conversion factor $c_f$ 
as a function of the WIMP mass $m_\chi$, assuming equilibrium. Derived from the calculations described in \cite{WSC}. The prominent peak around $50\unitx{GeV}$ is due to the resonant capture on Fe, the most abundant element in the Earth centre. At lower energies, also the presence of Si, Mg and O nuclei is relevant.}
\label{fig:conversion}
\end{figure}

For $t \gg (C_C C_A)^{-1/2}$ the value $\tanh^2\left(t \sqrt{C_C C_A}\right) \rightarrow 1$ and the capture and annihilation rates in the Earth reach equilibrium. 
The annihilation rate $\Gamma_A(t)$ then does not depend on $\langle\sigma_A v\rangle_{Earth}$ anymore and Eq.\ \ref{eq:A_r} simply becomes
\begin{equation}
\Gamma_{A,eq} = \frac{1}{2} C_C .
\end{equation}
In this case one can define a conversion factor $c_f$ between $\Gamma_{A,eq}$ and $\sigma^{SI}_p$ for a given $m_\chi$:
\begin{equation}
c_f=\frac{\Gamma_{A,eq}}{\sigma^{SI}_p}=\frac{\rho_{0.3}^{\chi}}{ 2 m_{\chi} \bar{v}_{270}} \sum\limits_{i} F_i^*(m_{\chi}).
\end{equation}
In Figure \ref{fig:conversion}, $c_f$ is plotted as a function of the WIMP mass.

Assuming in Eq.\ \ref{eq:C_A} a thermally averaged annihilation cross-section $\langle\sigma_A v\rangle_{Earth}$ equal to Eq.\ \ref{eq:sa}, i.e.\ the same as during the freeze out of WIMPs, the equilibrium condition is not generally satisfied.
Under this condition, $\tau \sim 10^{11}$~y, while the age of the Earth is $t^*\sim 4.5  \cdot 10^9$ y.  
In the case of non-equilibrium, the relationship between $\Gamma_A(t)$ and $\sigma^{SI}_p$ can be written as
\begin{equation}\label{eq:none}
\Gamma_A(t^*) = c_f \sigma^{SI}_p \tanh^2\left(t^* \sqrt{C_A 2 c_f \sigma^{SI}_p}\right) \underset{t^*\ll \tau}\longrightarrow \Gamma_A(t^*) \propto C_C^2 \cdot C_A.
\end{equation}
The annihilation rate (and thus the flux of neutrino-induced muons) depends quadratically on the capture factor and linearly on the annihilation factor.

The results presented in this paper assume spherically distributed DM with a Gaussian velocity distribution (standard halo model). The main astrophysical uncertainty that affects our result arises from the existence of a co-rotating structure made from materials accreted into the disc, known as dark disc \cite{M1}\cite{M2}. As reported in \cite{M3}, simulations show that the local density of the dark disc could range from a few percent up to $\sim 1.5$ times the density of the local dark matter halo, and with velocity distribution that varies for different scenarios.
The presence of a dark disc with high phase space density at low velocities enhances WIMP capture rates in the Earth up to a factor of 30 \cite{M4}. As the muon flux in a neutrino telescope depends on the annihilation rate (Eq.~\ref{eq:none}) and thus from the square of the capture rate, $C_C$, the presence of a dark disk could enhance our signal up to three orders of magnitude. For similar searches of DM signal from the Sun, the increase is an order of magnitude, as the muon flux depends on $C_C$. Direct searches are affected in a different way from this uncertainty, as scattering rate simply increases with the local density. In addition, as direct detection looks for energetic scattering of WIMPs, they are sensitive to WIMPs with high-velocity (less affected by the presence of the dark disc), while indirect detection techniques are sensitive to the low part of the velocity distribution.  Thus, limits expressed in the following sections are very conservative with respect to the presence of a dark disk.

\section{Signal and background modelling in the ANTARES neutrino telescope}\label{ch_3}
The ANTARES neutrino telescope is a deep sea water Cherenkov detector, located 40 km offshore from Toulon \cite{ANTA18}. 
The detector is anchored at the seabed, at a depth of about $2475\unitx{m}$. It consists of 885 optical modules (OMs) \cite{OMpaper}. Each optical module houses one 10'' photomultiplier tube looking downward with an angle of $45^\circ$. The OMs are arranged in storeys, with 3 OMs per storey. 
The storeys are connected by a flexible cable \cite{alignment} and form lines, with 25 storeys per line and 12 lines in the detector\footnote{On one of the lines only 20 storeys hold optical modules.}. The vertical distance between storeys is $14.5\unitx{m}$, the length of a line is about $450\unitx{m}$, and the horizontal spacing between the lines is $60 - 70\unitx{m}$.\par
The detection principle of ANTARES is based on the observation of Cherenkov photons, induced by charged secondary particles produced in interactions of neutrinos around or inside the instrumented volume.

\begin{figure}
\centering
\begin{minipage}{.5\textwidth}
  \centering
  \includegraphics[width=1\linewidth]{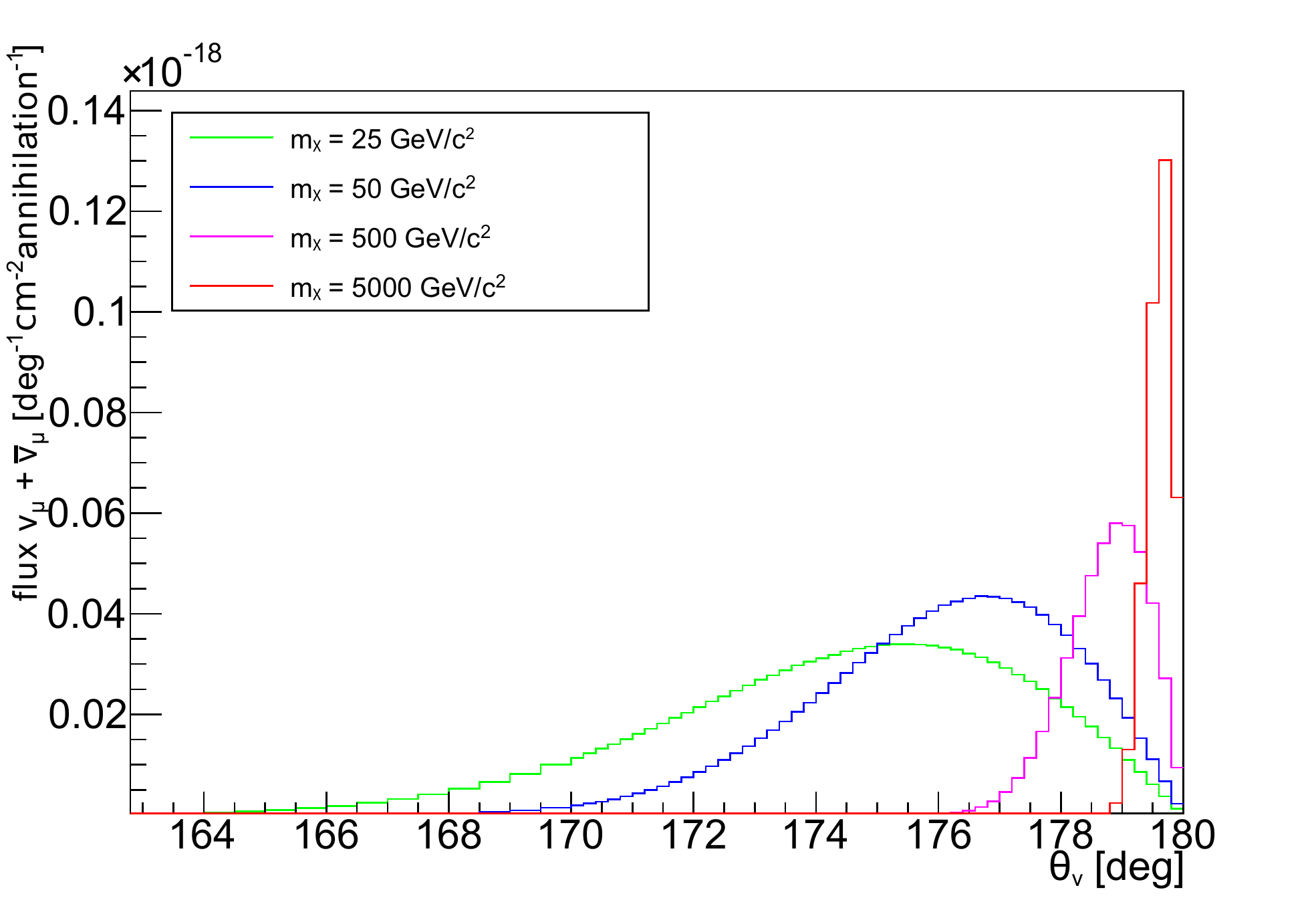}
  \label{fig:test1}
\end{minipage}%
\begin{minipage}{.5\textwidth}
  \centering
  \includegraphics[width=1\linewidth]{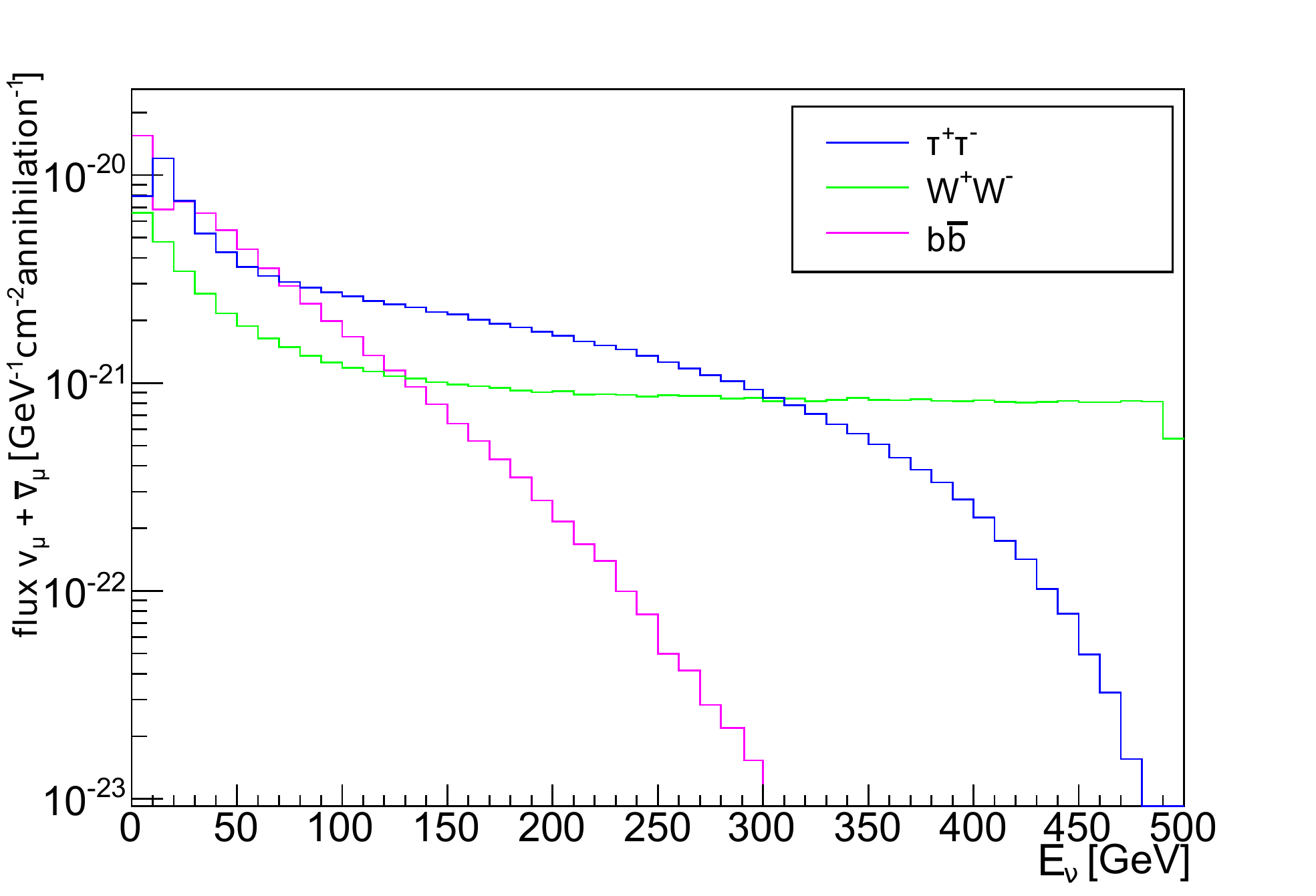}
  \label{fig:test2}
\end{minipage}
\caption{Left: Zenith angle distribution of the differential $\nu_{\mu}$ + $\overline{\nu}_{\mu}$ flux per WIMP pair annihilation through the $\tau^{+}\tau^{-}$ channel for different WIMP masses at the surface of the Earth. Here $\theta_\nu=180^\circ$ corresponds to vertically upward going neutrinos. Right: Energy spectrum of the differential $\nu_{\mu}$ + $\overline{\nu}_{\mu}$ flux per WIMP annihilation (in $\tau^{+}\tau^{-}$, $W^+W^-$ and $b\overline{b}$ pairs) for  $m_\chi=500\unitx{GeV}$ at the surface of the Earth. Simulation is done with WimpSim \cite{WS2}\cite{WS}.}
\label{fig:test1w}
\end{figure}

The study presented in this paper is based on muons from charged current $\nu_{\mu}+ \overline \nu_\mu $ interactions.
WIMP masses between $25\unitx{GeV}$ and $1000\unitx{GeV}$ are considered. Concerning the annihilation channels, several cases have been studied: the $b\overline{b}$ channel (which gives a soft spectrum of neutrinos), the $\tau^{+}\tau^{-}$ or  $W^+W^-$ channels (which give a hard spectrum of neutrinos) and the monochrome, non-SUSY, $\nu_{\mu}\bar{\nu}_{\mu}$ channel. In each case, a $100\%$ branching fraction is assumed.
The neutrino flux from DM annihilations in the Earth is simulated and propagated to the detector using WimpSim \cite{WS2}\cite{WS}.
The code includes neutrino interactions and neutrino oscillations in a complete three-flavour treatment (the values $\theta_{12} = 33.58^\circ$,  $\theta_{13} = 9.12^\circ$, $\theta_{23} = 40.40^\circ$, $\delta_{CP}  = 0$, $\Delta m^2_{21} = 7.58 \cdot 10^{-5}\unitx{eV^2}$, $\Delta m^2_{31} = 2.35\unitx{eV^2}$ are used).
The expected neutrino fluxes in ANTARES for some cases are shown in Figure \ref{fig:test1w}, as a function of zenith angle and energy.
It can be noticed that most of the signal is expected from around the nadir.

The primary sources of background in this analysis consist of muons and neutrinos which have their origin in interactions of cosmic rays with the atmosphere of the Earth. 
The atmospheric muons are simulated with the MUPAGE \cite{MUPAGE} package, which uses parametric formulae of the fluxes of muon bundles \cite{MUPAGE-muonbundles1}.
For the background from atmospheric neutrinos, only charged-current $\nu_{\mu}$ + $\overline{\nu}_{\mu}$ interactions contribute significantly. For the conventional neutrino flux, the parameterisation of \cite{bartol} is used with a prompt contribution according to \cite{enberg}. 
For atmospheric neutrinos, oscillations are taken into account in a two-flavour scenario (using the same values for $\Delta m^2_{31}$ and $\theta_{23}$ as reported before).

High quality data runs, defined according to environmental and data taking conditions, are selected for this work (analogously to \cite{ANTA38}). A detailed Monte Carlo simulation is available for each data acquisition run \cite{margi}.

Two different methods of event reconstruction are employed. The first is a fast muon track reconstruction algorithm called Qfit \cite{Qfit}. It is based on the minimisation of a $\chi^2$-like quality function which uses the differences between expected and measured arrival times of the photons at the OMs (time residuals). Qfit is well suited for low energy events, which are often only detected by the OMs of a single line.\par
The second method is a more sophisticated muon track reconstruction algorithm called $\Lambda$fit \cite{ANTA38}. It is based on the maximisation of a likelihood function which again uses time residuals. $\Lambda$fit is better suited for higher energetic events.

\section{Event selection and reconstruction}\label{ch_4}

For this paper, data collected with the ANTARES neutrino telescope from 2007 to 2012, with a livetime of 1191 days have been analyzed.
As shown in Figure \ref{fig:test1w}, the flux from dark matter annihilations is restricted to certain (narrow) cones from the centre of the Earth and to a certain energy range. 
As the background increases quadratically with the search cone, the size of the cone which maximizes the signal-to-background ratio is determined for different WIMP mass intervals.
Furthermore, the energy range in which neutrino candidates are searched for is determined as a function of the WIMP mass and annihilation channel.
The event selection criteria are therefore chosen in a way that they restrict the minimal zenith angle and maximal energy of the events.
The former is done by selecting events which are reconstructed by either Qfit or $\Lambda$fit with a high zenith angle $\theta_Q$ and $\theta_\Lambda$, respectively.  
The latter is estimated by evaluating the muon range, as described below. By discarding muons for which the estimated range exceeds a certain threshold, events which are not likely to be signal are discarded.

To reduce the background from atmospheric muons, an additional requirement strictly selects reconstructed zenith angles close to the vertical direction.
The criterion imposes that the hits are compatible with the expected signature from muons arriving from a direction close to the vertical. For each photomultiplier, a hit is defined by the arrival time of the first photon and the number of photons (amplitude).
For those events the vertical muon range $R_{proj}$  can be calculated as the distance between the highest ($N_H$) and lowest ($N_L$) storey which registered a hit: $R_{proj} = (N_H-N_L)\times 14.5\unitx{m}$. Near the nadir, $R_{proj}$ approaches the muon range, that is a proxy for the muon energy, and consequently a proxy of the parent neutrino energy. Due to limits of the detector geometry and reconstruction method, $3 \le N_H-N_L \le 24$. $R_{proj}$ is used to select low energetic muons by requiring $R_{proj}$ to be less or equal to a given cut value. 
\begin{figure}[tbh]
\centering
\includegraphics[angle=0,width=12cm] {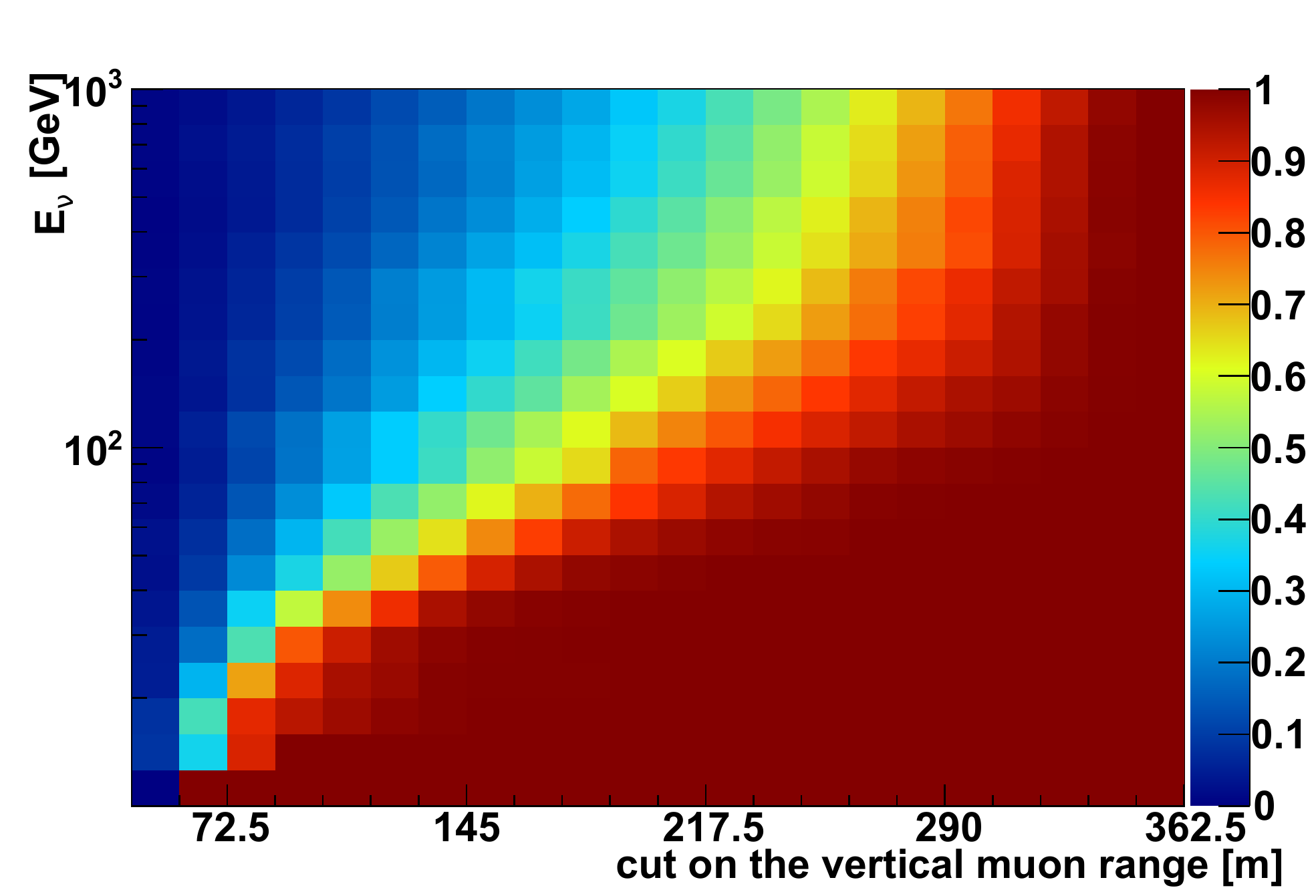}
\caption{The simulated fraction of remaining events (coloured scale) as a function of the true energy of the neutrino and the cut on the reconstructed vertical muon range. Only events which are reconstructed with a zenith angle of at least $175^\circ$ are considered. The coloured scale can be understood as the probability that a muon, originating from a neutrino with a certain energy (y-axis), is accepted given a certain cut (x-axis) on its reconstructed vertical range.}
\label{fig:E_ZAVs1}
\end{figure}

Figure \ref{fig:E_ZAVs1} shows the fraction of surviving events as a function of the true neutrino energy and of the maximum estimated muon range. 
For example, if one discards all events with a reconstructed vertical muon range $>217.5\unitx{m}$, more than $99\%$ of the vertical neutrinos of energy below $40\unitx{GeV}$ would be accepted.
Using this kind of information, the signal acceptance for a given WIMP mass and annihilation channel is estimated by folding the energy-dependent fraction of surviving events with the energy-dependent flux of the signal neutrinos.

\begin{figure}
\centering
  \includegraphics[width=0.6\linewidth]{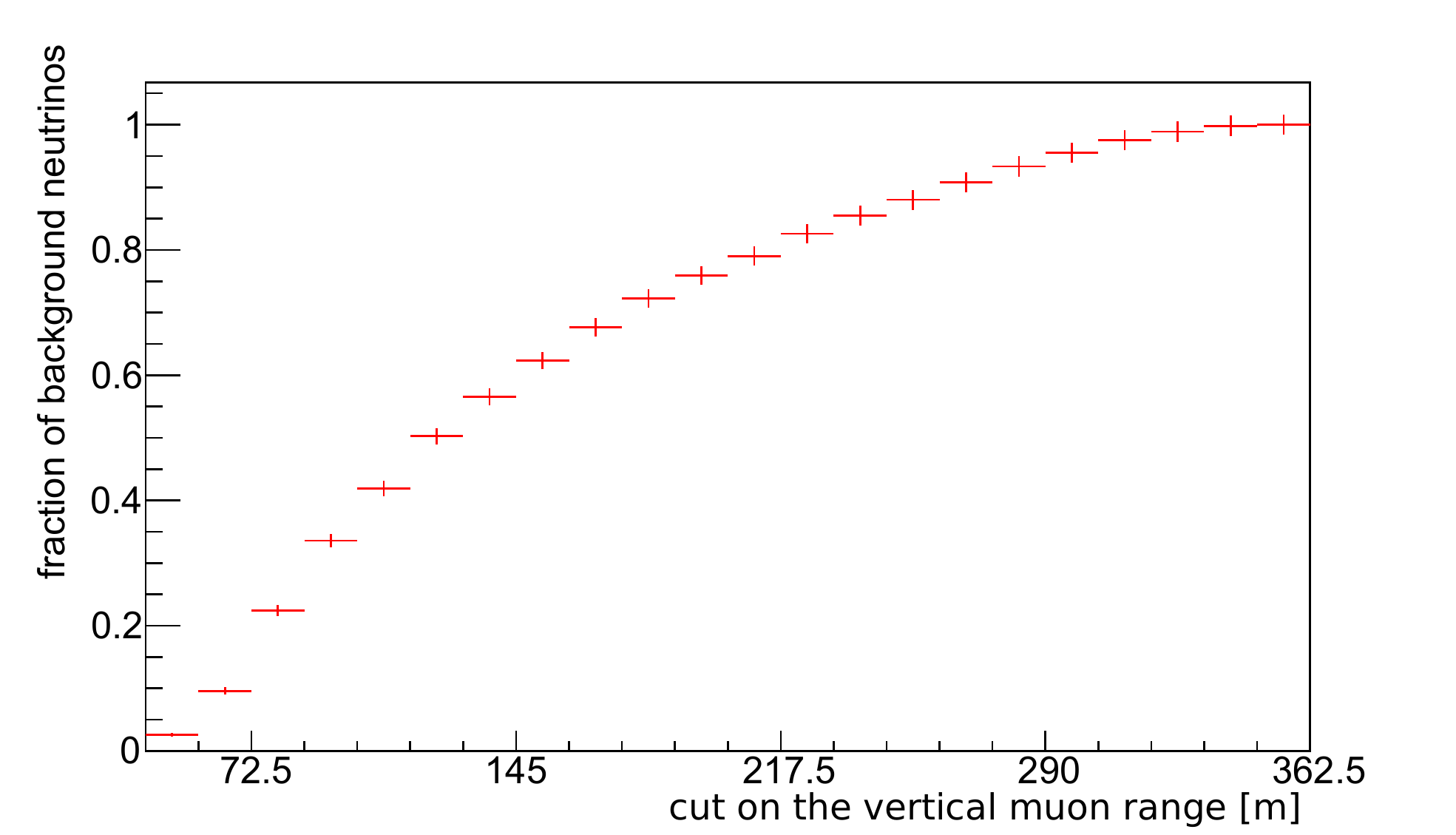}
\caption{The simulated fraction of background atmospheric neutrinos remaining after cuts on the reconstructed vertical muon range. Only events which are reconstructed with a zenith angle of at least $175^\circ$ are considered.}
\label{fig:E_ZAVs2}
\end{figure}

The fraction of remaining background due to atmospheric neutrinos as a function of the reconstructed vertical muon range is shown in Figure \ref{fig:E_ZAVs2}. 
The cut on the reconstructed vertical muon range is particularly efficient to enhance the signal-to-noise background for low WIMP masses and annihilation channels with softer neutrino spectra. 
For instance, requiring $R_{proj} \leq 217.5\unitx{m}$ (the same value as in the example above), rejects more than $20\%$ of the neutrino background.
For small WIMP masses and soft annihilation channels, an additional selection criterion is used by requiring that the lowest storey of the detector did not register a hit.
This reduces the background due to those atmospheric muons passing outside the instrumented volume and for which a radiative process induces a large energy loss just below it.
Such events can produce upward going charged mesons \cite{MACRObacksca} and mimic vertically upward going muons entering the detector from below.

\begin{figure}
\begin{tabular}{cc}
  \includegraphics[width=.5\textwidth]{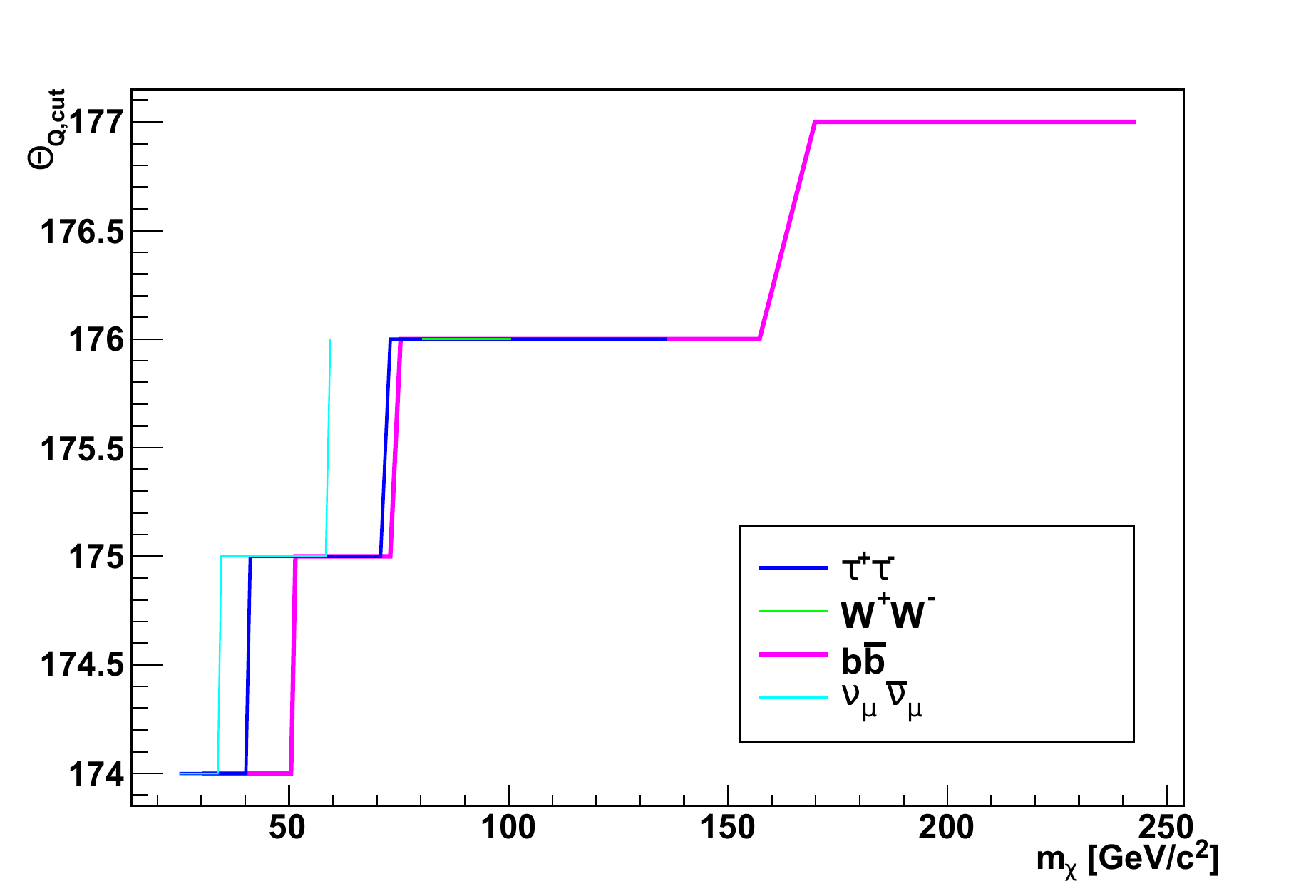} &   \includegraphics[width=.5\textwidth]{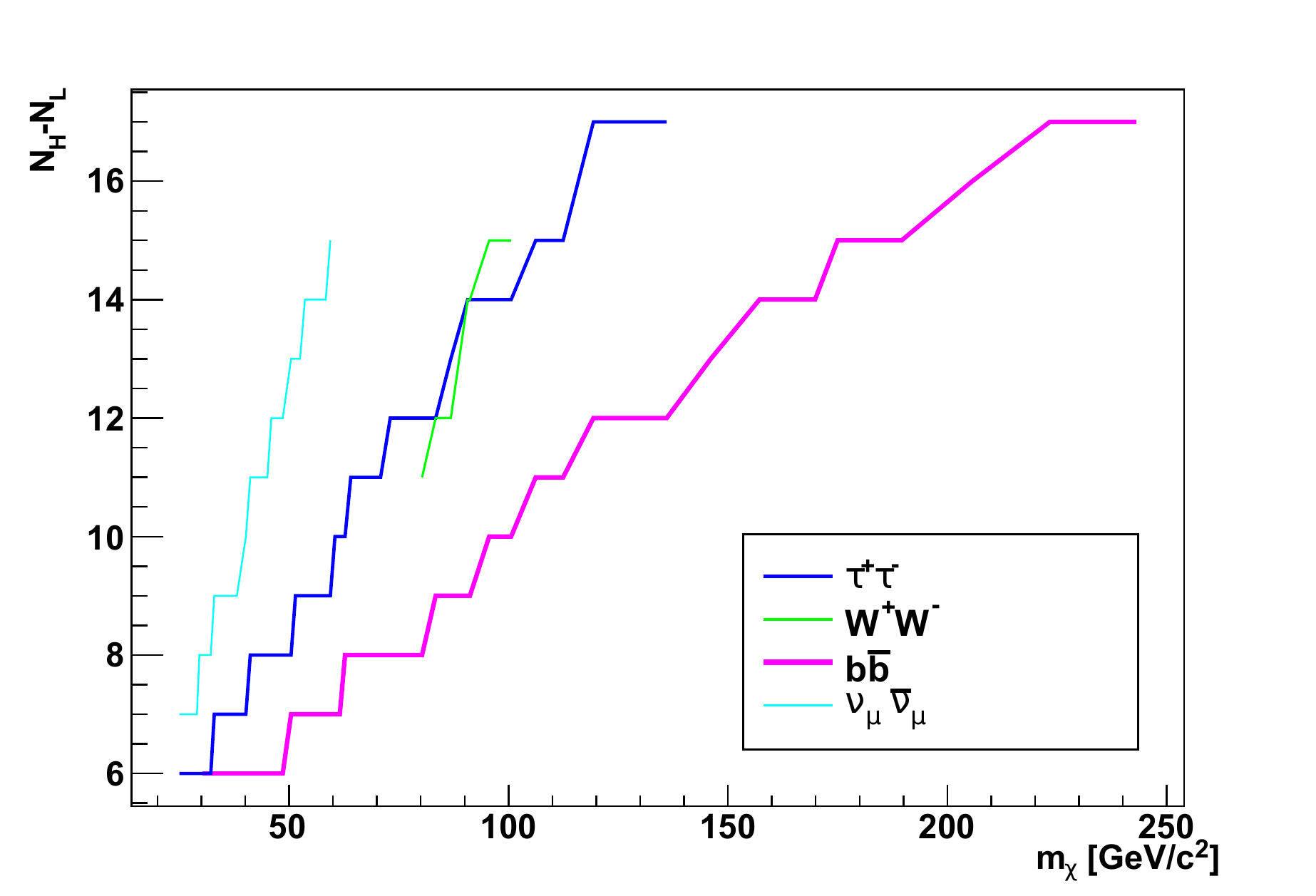}  \\
  \includegraphics[width=.5\textwidth]{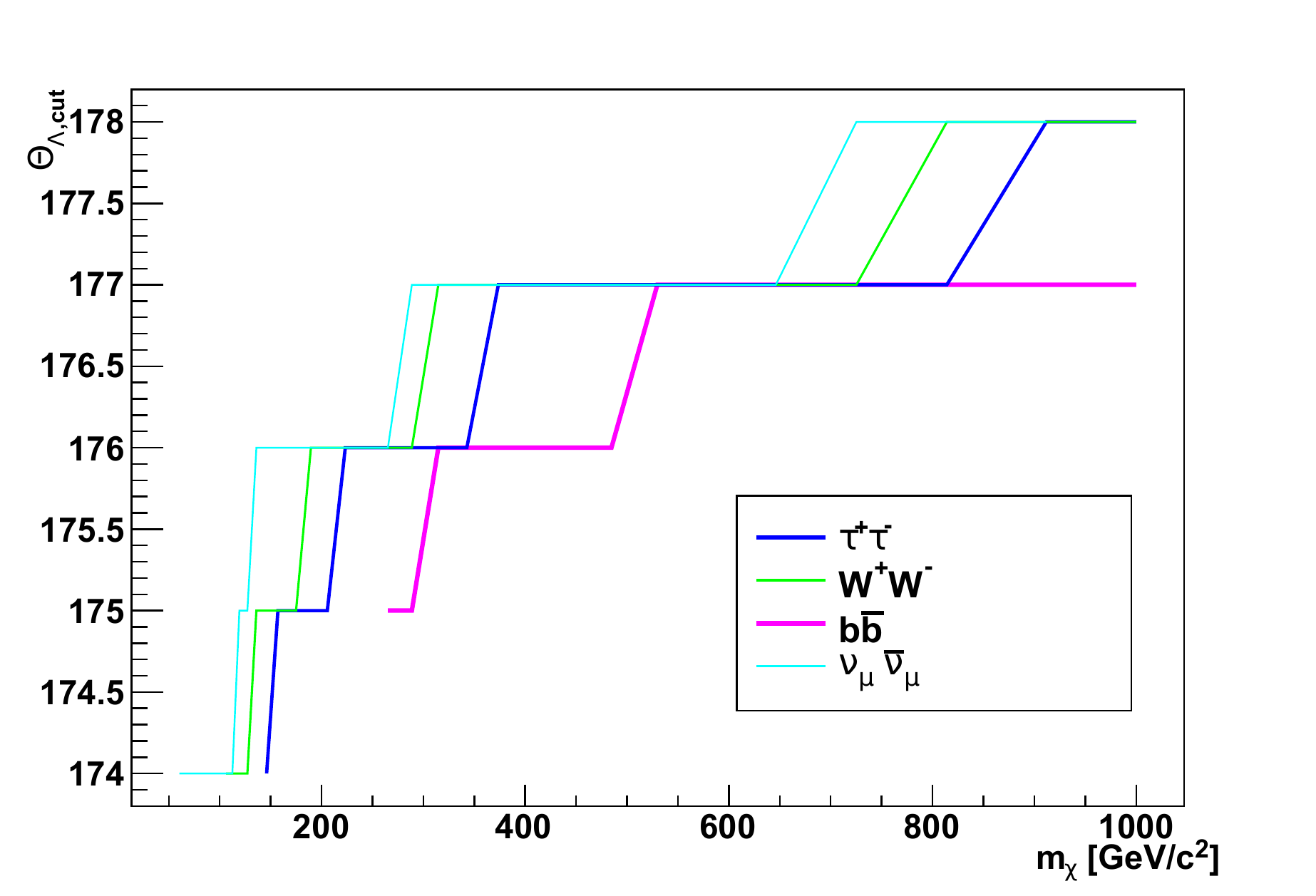} &   \includegraphics[width=.5\textwidth]{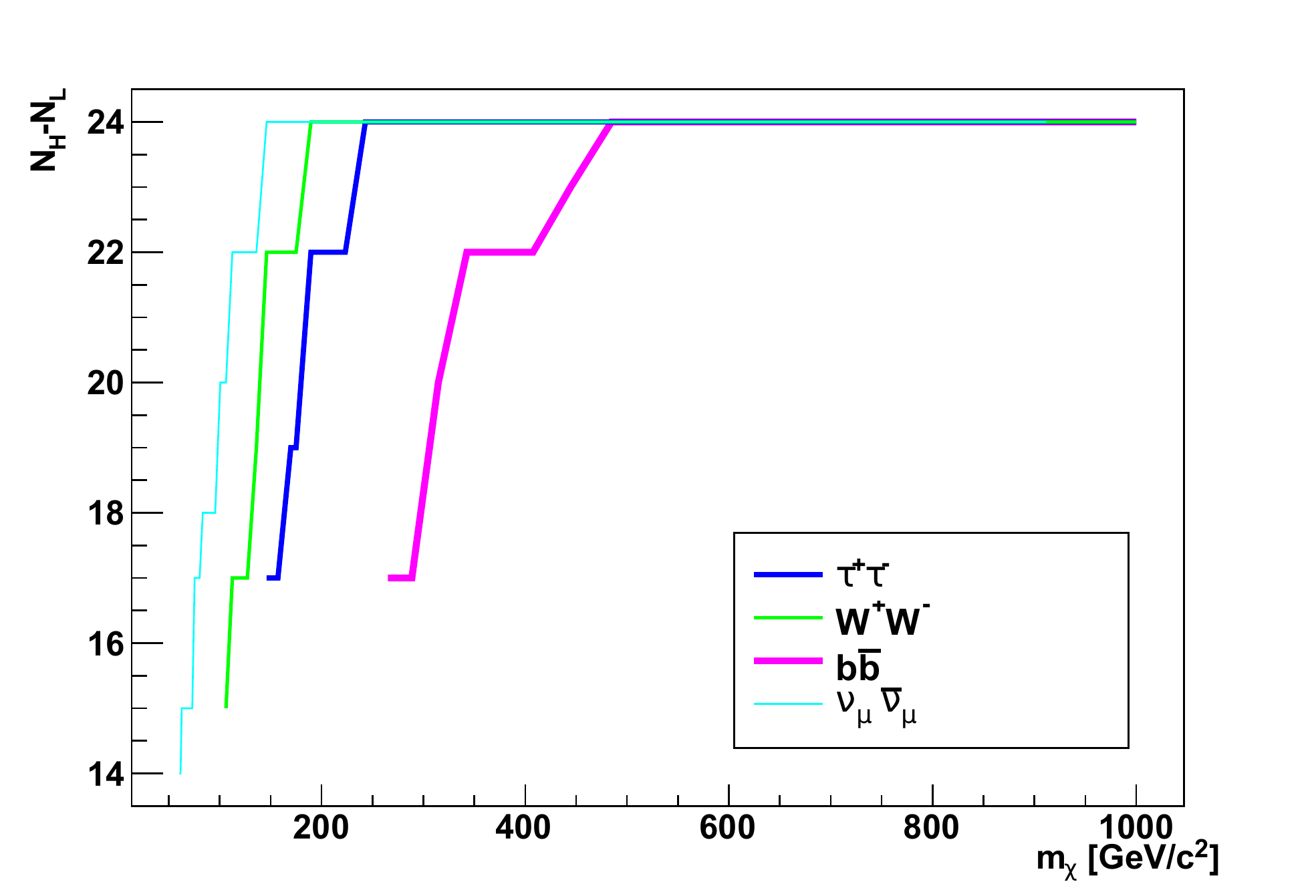} \\
\end{tabular}
\caption{Values of the zenith angle cut (left panels) and of the values of $N_H-N_L$ (right panels) that optimize the selection criteria as a function of the WIMP mass. The two top (bottom) panels refer to Qfit ($\Lambda$Fit).}
     \label{opt_cuts_1}
\end{figure}

A final cut requires that the uppermost storeys of the detector did not register any hit. This removes high-energy passing muons.
As expected (compare with Figure \ref{fig:test1w}), harder zenith angle cuts and looser energy cuts are more efficient for higher WIMP masses as presented in Figure \ref{opt_cuts_1}.
The event selection criteria have been chosen with the approach for unbiased cut selection for optimal upper limits \cite{MRF}. The cut parameters have been tuned individually for each annihilation channel and several WIMP masses in the considered mass range. 
The WIMP annihilation rate $\Gamma_A(t)$ has been used as scaling parameter of the source flux.

\section{Results}\label{ch_5}

In most ANTARES analyses, a signal-free control region of the same detector acceptance can be defined (see for instance \cite{ANTA52}). This is not possible for this particular analysis. Consequently, the background estimate in the signal region can only be derived from simulations. 
The reliability of the Monte Carlo is verified by comparing the simulations and data in a test region that presents almost the same detector response as the signal region, a similar background, but contains only a minimal residual fraction of the signal. 
Based on the results of WimpSim, this region corresponds to the interval of reconstructed zenith angles between $160^\circ$ and $170^\circ$.

\begin{figure}
\begin{tabular}{cc}
  \includegraphics[width=.5\textwidth]{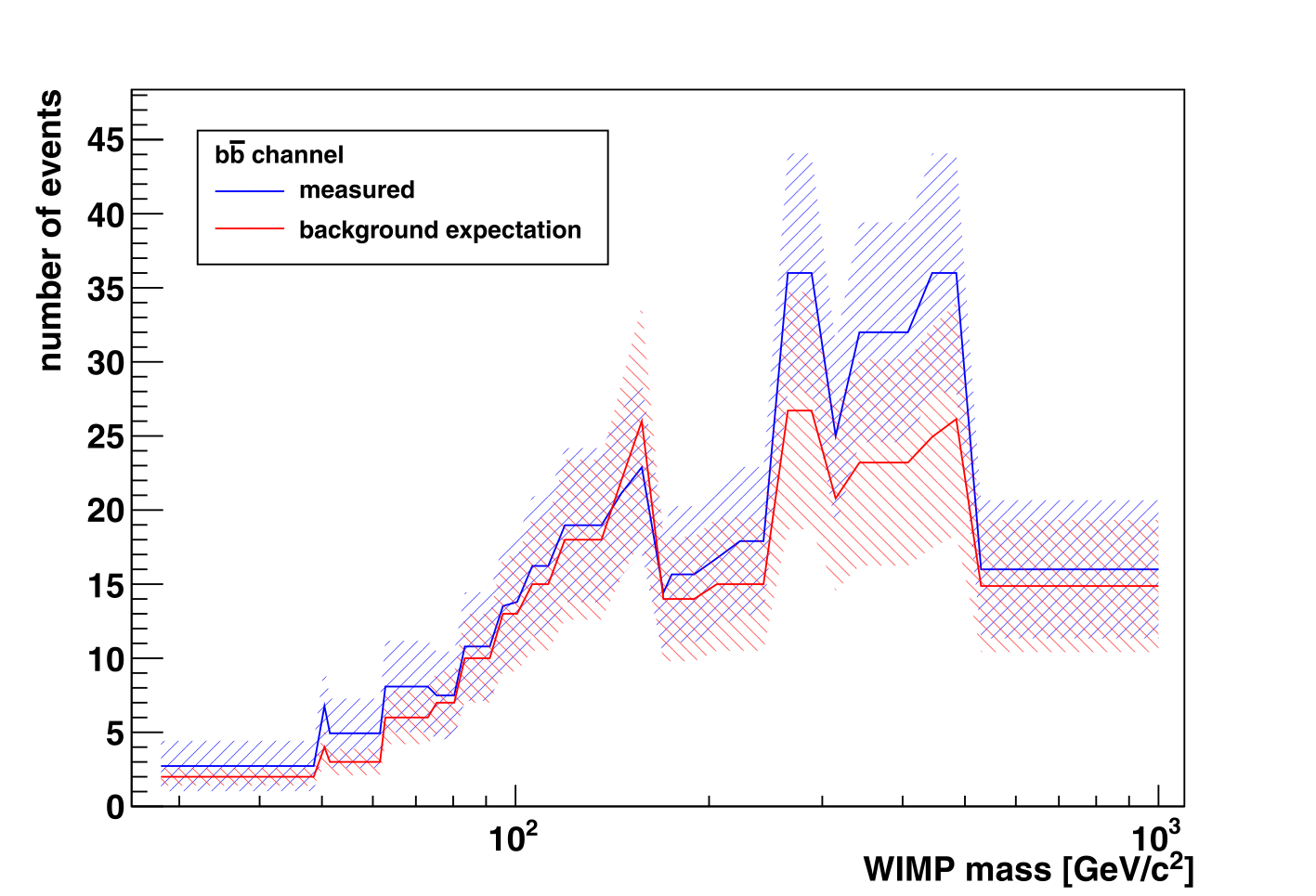} &   \includegraphics[width=.5\textwidth]{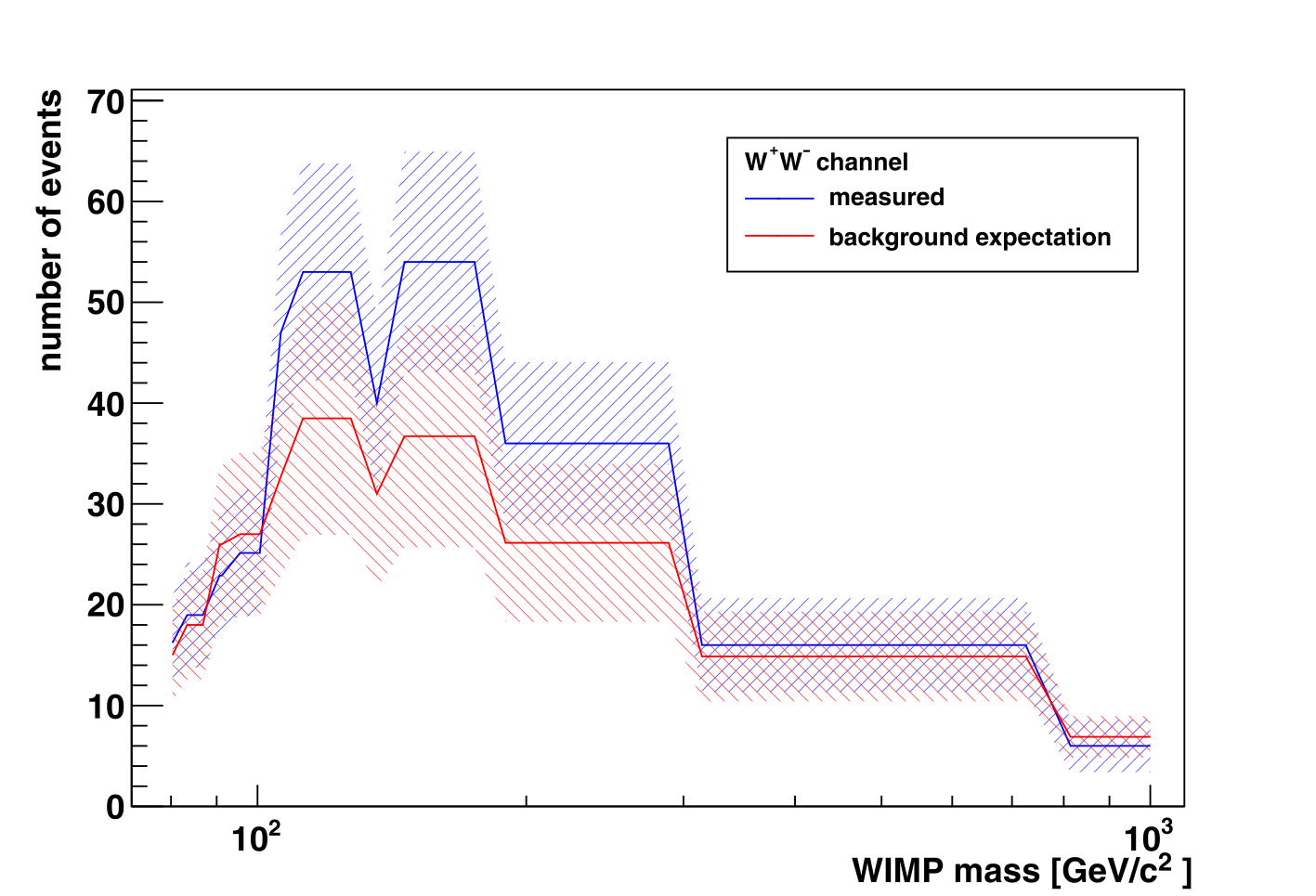} \\
\scriptsize(a)  channel = $b\overline{b}$ & \scriptsize(b) channel = $W^+W^-$ \\
  \includegraphics[width=.5\textwidth]{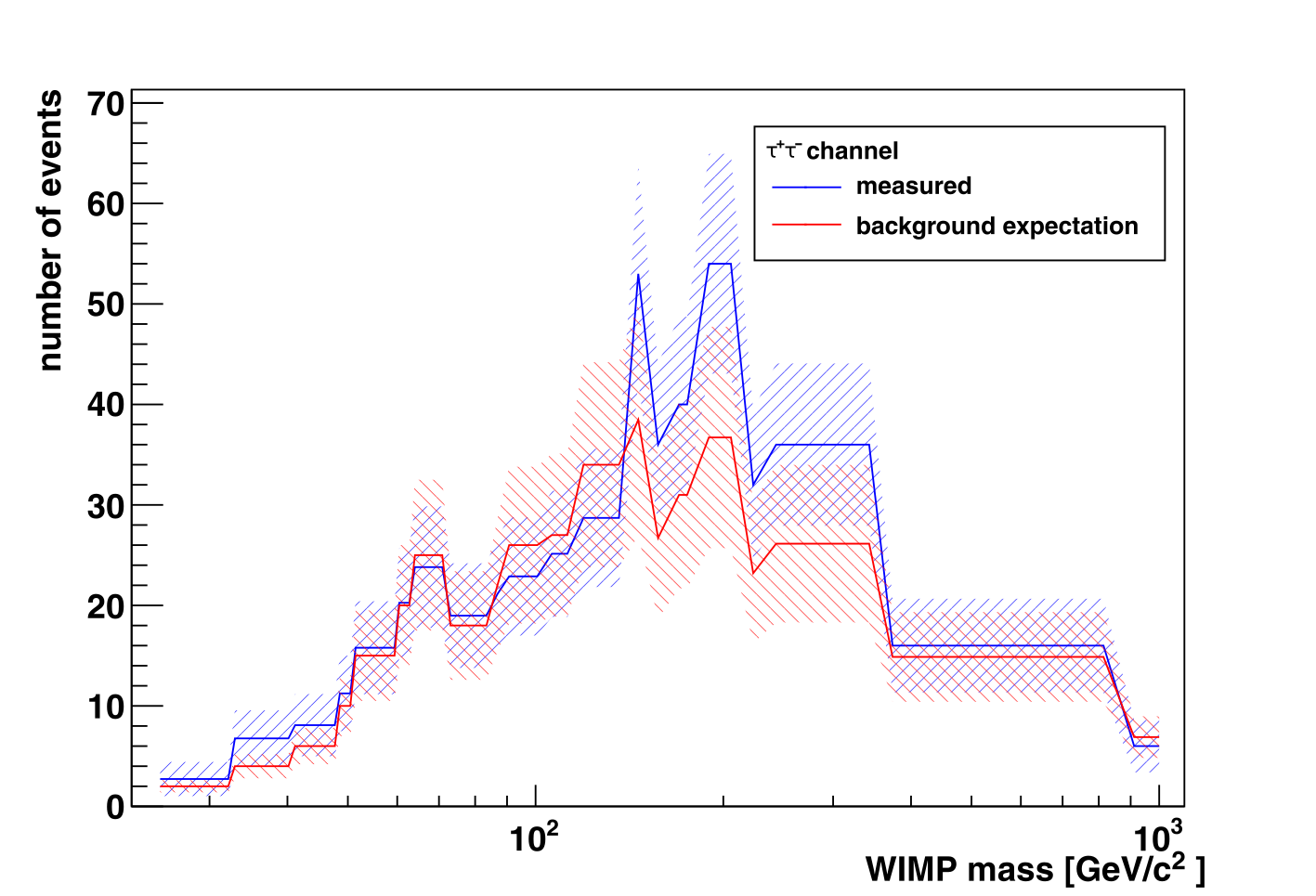} &   \includegraphics[width=.5\textwidth]{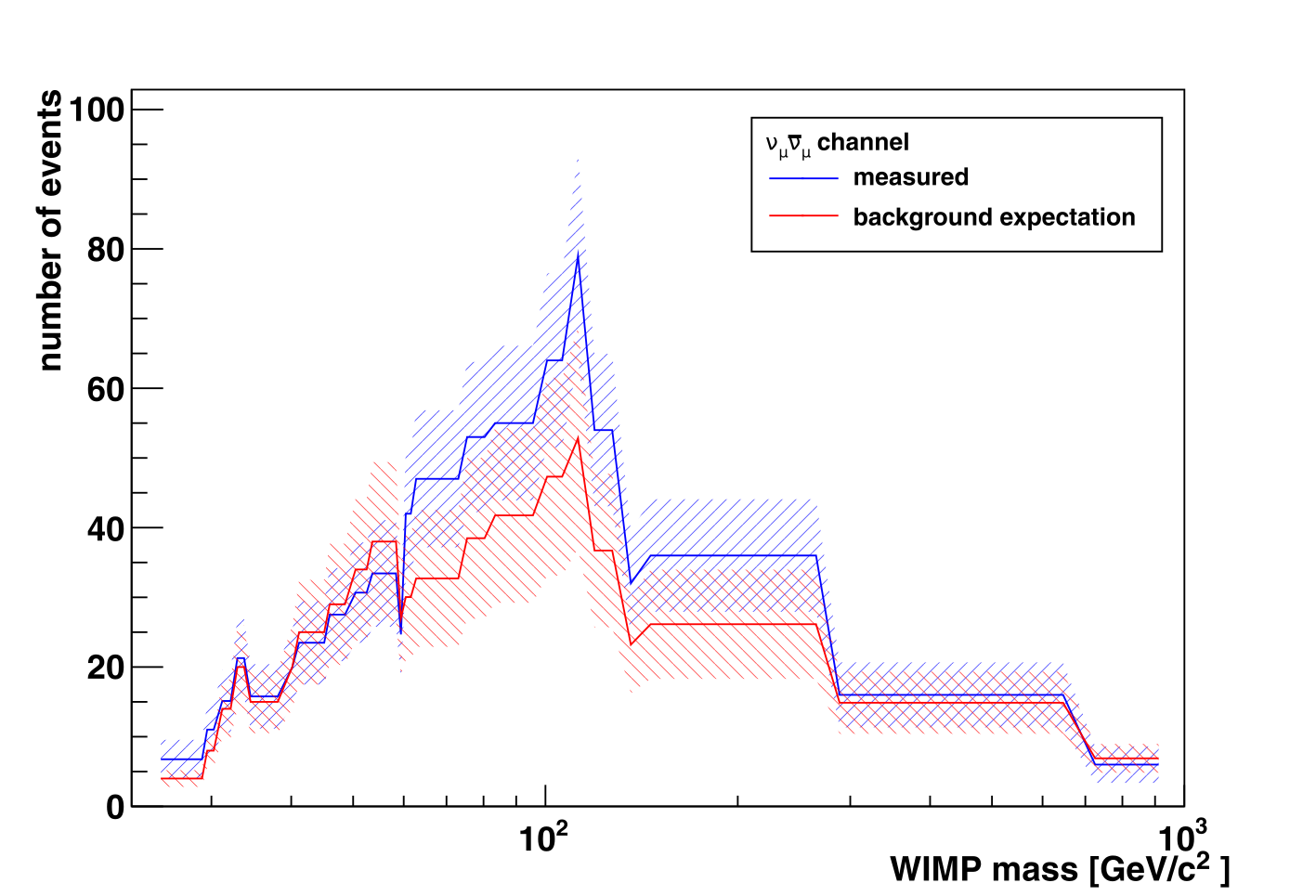} \\
\scriptsize(c) channel =  $\tau^{+}\tau^{-}$ &\ \scriptsize(d) channel = $\nu_{\mu}\bar{\nu}_{\mu}$ \\
\end{tabular}
\caption{Number of background events (red) and observed events (blue) as a function of the WIMP mass for the four considered annihilation channels. In both cases, the shaded regions represent the quadratic sum of the statistical and systematic uncertainties.}
\label{fig:events}
\end{figure}

After the cuts, no simulated atmospheric muons were survive while atmospheric neutrinos represent the irreducible background. The residual number of atmospheric neutrinos after the application of cuts defined for the four considered WIMP annihilation channels as a function of $m_\chi$ are reported in Figure \ref{fig:events}. 
The same plot shows the number of selected events. 
When comparing the signal to the background, some excess of events is observed for certain channels and $m_\chi$. The uncertainty used in the computation of the upper limits are due to the uncertainty on the background estimation and on the detector acceptance. The latter affects both the number of signal and background events. According to \cite{nu-uncertainty}, a systematic uncertainty of $30\%$ on the atmospheric neutrino flux can be assumed. 
This is in agreement with the ANTARES measurement of the atmospheric muon neutrino spectrum \cite{ant34}, in which the overall normalization factor for atmospheric muon neutrinos is increased by $\sim25\%$ to match data. 
Concerning the uncertainties related to the detector acceptance and efficiency, a systematic uncertainty of $15\%$ is assumed, following \cite{ANTA28}\cite{ANTA15}. This overall $15\%$ effect is mainly due to the uncertainties on water properties, on the uncertainty on the optical module (OM) angular acceptance and on OM efficiencies. The effect on signal and background was derived using dedicated Monte Carlo simulations with modified water and OM parameters. To be conservative, in the upper limit computation only the increases of the background level were considered. Finally, no atmospheric muon survives the cuts and their contribution to the background is assumed to be equal to zero in all considered channels. Because the number of simulated atmospheric muons corresponds only to 1/3 of the trigger rate,  the uncertainty associated with this non contribution to the background corresponds to  6.9 events, i.e. three times the $90\%$ C.L. upper limit assuming Poisson distribution. This is the most conservative approach for setting upper limits.

When considering the above uncertainties, no significant excess of events is observed.
From this, a $90\%$ C.L. upper limit  in the number of events $\mu_{90\%,R}$ is calculated using the TRolke module from ROOT \cite{Trolke}. This class computes confidence intervals for the rate of a Poisson process with the considered background and efficiency uncertainties using a fully frequentist approach with the profile likelihood method \cite{rolke}.

\begin{figure}
  \centering
  \includegraphics[angle=0,width=9cm]{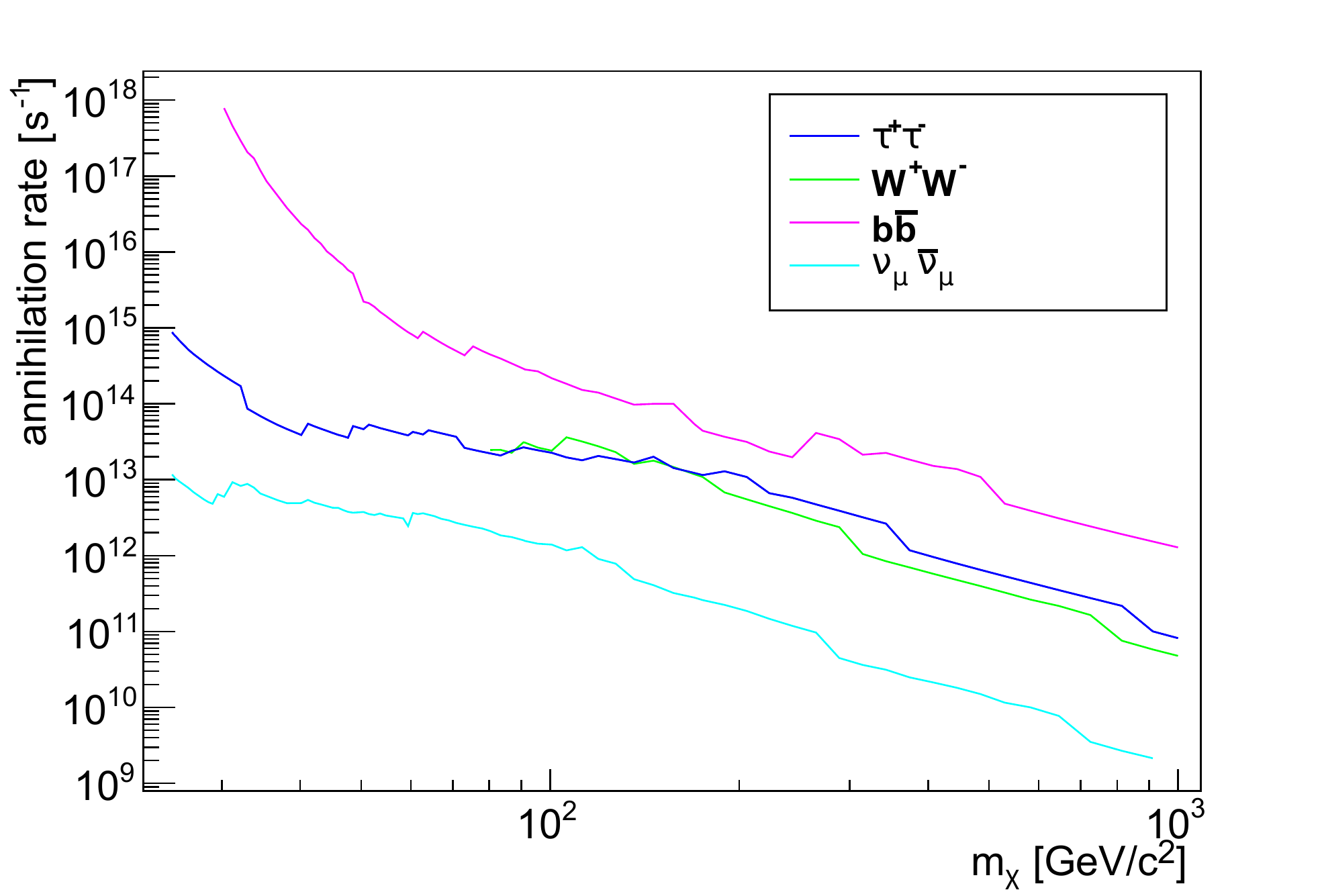}
  \caption{$90\%$ C.L. upper limits on $\Gamma_A$ as a function of the WIMP mass. For each channel, the WIMP pair annihilates to $100\%$ into either $\tau^{+}\tau^{-}$, $W^+W^-$, $b\overline{b}$ or $\nu_{\mu}\bar{\nu}_{\mu}$. The lowest WIMP masses shown are at $25\unitx{GeV}$.}
  \label{fig:arlimit}
\end{figure}

From $\mu_{90\%,R}$, the $90\%$ C.L. upper limits on the WIMP annihilation rate $\Gamma_{A,90\%}$ in the Earth is calculated as:
\begin{equation}
 \Gamma_{A,90\%}=\frac{\mu_{90\%,R}}{n_s}\cdot \Gamma_{0}
\end{equation}
Here $n_s$ is the number of expected signal events for this experiment in the 2007 -- 2012 data according to simulations, assuming an annihilation rate  $\Gamma_A \equiv \Gamma_{0} = 1 \unitx{annihilation/s}$. 
The values of $\Gamma_{A,90\%}$ as a function of the WIMP mass and annihilation channel are shown in Figure \ref{fig:arlimit}.

\begin{figure}
\centering
  \includegraphics[width=0.8\linewidth]{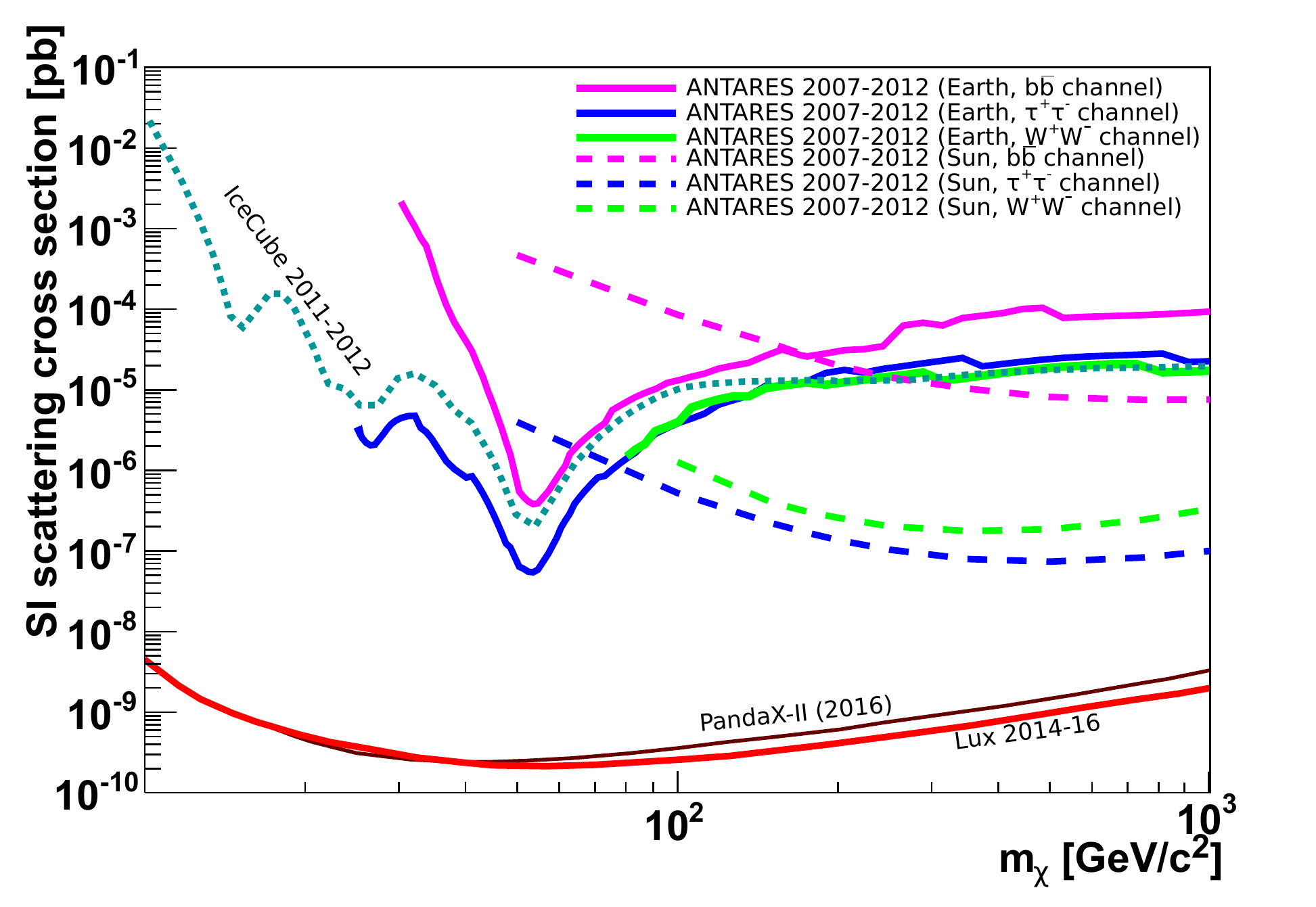}
\caption{$90\%$ C.L. upper limits on $\sigma^{SI}_p$ as a function of the WIMP mass for ANTARES 2007 -- 2012 (Earth) and ANTARES 2007 -- 2012 (Sun)  \cite{antares_dmsun}, assuming $\langle\sigma_A v\rangle_{Earth} = 3 \cdot 10^{-26}\unitx{cm^3 s^{-1}}$ and WIMP pair annihilation to $100\%$ into either $\tau^{+}\tau^{-}$ (blue),  $W^+W^-$ (green) or $b\overline{b}$ (purple). Also Shown are the results IceCube-79 2011 -- 2012 (Earth, $\tau^{+}\tau^{-}$ channel for WIMP masses $< 80.4\unitx{GeV}$ and $W^+W^-$ channel for WIMP masses $\geq 80.4\unitx{GeV}$) \cite{icecubeearth}, PandaX-II (2016) \cite{PandaX} and LUX \cite{lux_results}. The prominent dip at around 50 GeV is a common feature for all indirect searches from the centre of the Earth, see Figure \ref{fig:conversion}.}
\label{fig:limits_s}
\end{figure}

From the limits on the annihilation rate, limits on $\sigma^{SI}_p$ are derived as described in section \ref{ch_2} assuming the natural scale (see Eq.\ \ref{eq:sa}) for the thermally averaged annihilation cross-section times velocity. 
In this case, equilibrium is not reached and only SUSY-allowed annihilation channels are considered. 
The upper limits derived with this search on the spin-independent cross-section $\sigma^{SI}_p$ as a function of the WIMP mass $m_\chi$ are shown in Figure \ref{fig:limits_s}.
They are compared with limits from other indirect and direct dark matter searches.
The results presented here set the most stringent limits for indirect searches in the mass interval from about $40$ to $70\unitx{GeV}$, where the WIMP capture factor is enhanced due to the heavy composition of the Earth. 

\begin{figure}
  \includegraphics[width=1\textwidth]{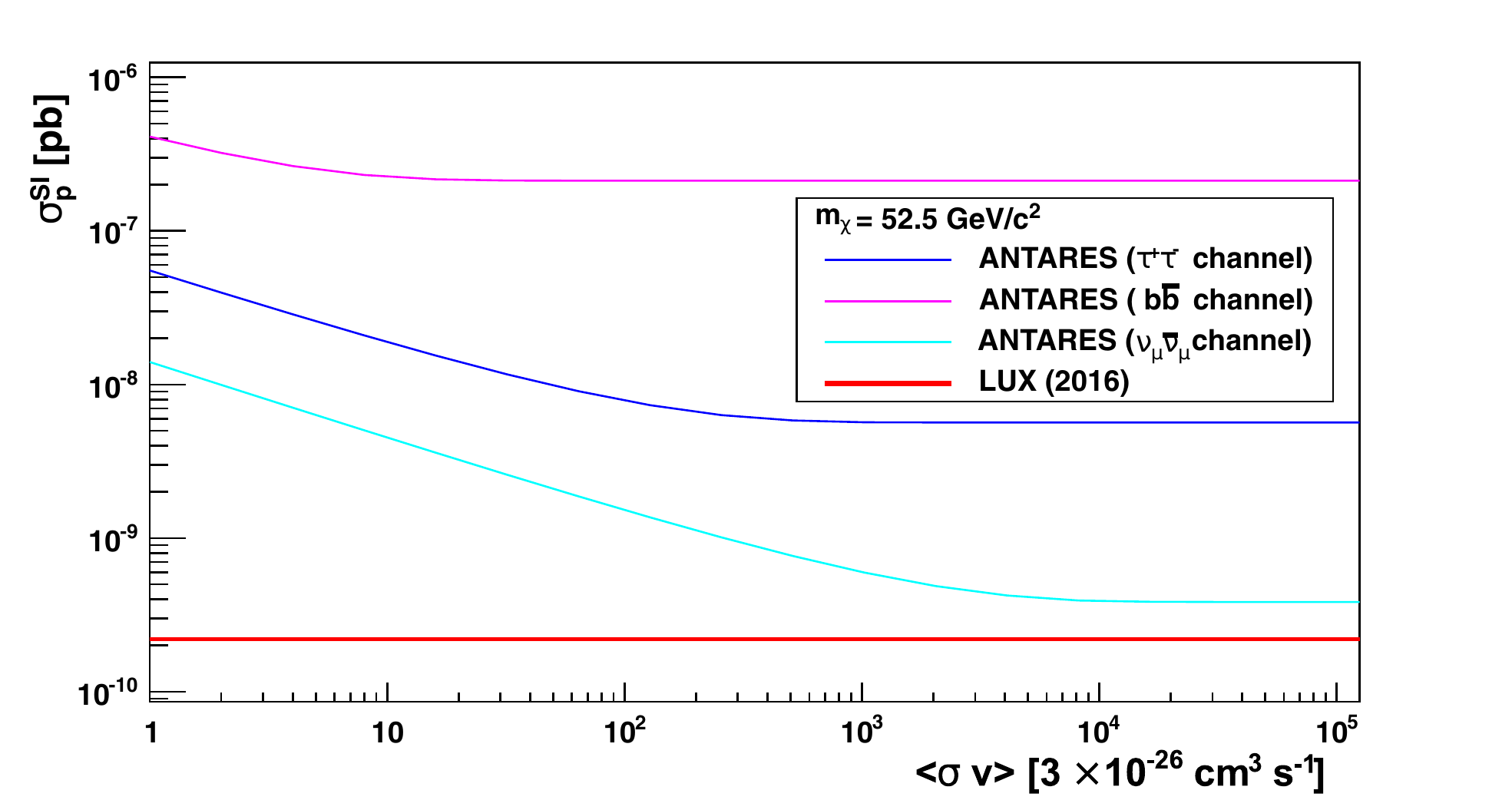}  
\caption{ANTARES 2007 - 2012 $90\%$ C.L. upper limits on $\sigma^{SI}_p$ as a function of $\langle\sigma_A v\rangle_{Earth}$ (in units of $3 \cdot 10^{-26}$ cm$^3$ s$^{-1}$) for $m_\chi=52.5$ GeV. For this WIMP mass, the capture of WIMPs in the Earth is strongly enhanced due to the presence of the iron resonance. The WIMP pair annihilation is assumed to be $100\%$ into either $\tau^{+}\tau^{-}$,  $\nu_\mu \overline\nu_\mu$ or $b\overline{b}$. For comparison, the LUX \cite{lux_results} limit for the same $m_\chi$ is shown.}
     \label{fig:limits_s_b}
\end{figure}

In addition, a scenario where $\langle\sigma_A v\rangle_{Earth}$ is enhanced compared to the value during the freeze out of WIMPs has also been considered.
In this case, the non-SUSY $\nu_\mu \overline{\nu}_\mu$ annihilation channel is also considered.  
The upper limits on $\sigma^{SI}_p$ as a function of $\langle\sigma_A v\rangle_{Earth}$ are shown in Figure \ref{fig:limits_s_b}, assuming $m_\chi=52.5\unitx{GeV}$. This corresponds to a mass where the capture of the WIMPs in the Earth is strongly enhanced due to the presence of the iron resonance (Figure \ref{fig:conversion}).

\section{Conclusion}\label{ch_6}

In this paper, the results of a search for neutrinos from dark matter annihilation in the centre of the Earth using data taken with the ANTARES neutrino telescope from 2007 to 2012 (corresponding to a lifetime of 1191 days) have been presented. 
The number of neutrinos observed from the direction of the centre of the Earth is compatible with the background expectation from atmospheric events. 
Assuming the natural scale for $\langle\sigma_A v\rangle$, the $90\%$ C.L. upper limits on the WIMP self-annihilation rates have been set as a function of the WIMP mass. 
WIMP pair annihilation into either $\tau^{+}\tau^{-}$,  $W^+W^-$, $b\overline{b}$ or (non-SUSY) $\nu_{\mu}\bar{\nu}_{\mu}$ channels have been considered. 
These are translated into limits on the spin independent scattering cross-section of WIMPs off protons. 
A scenario where the annihilation cross-section for dark matter in the Earth is enhanced compared to the value during the freeze out of WIMPs has also been considered. The limits derived by this search are competitive with other types of indirect dark matter searches.
In particular, the results presented here set the most stringent limits for indirect searches in the mass interval from about $40$ to $70\unitx{GeV}$.

\section{Acknowledgements}\label{ch_6}
The authors acknowledge the financial support of the funding agencies:
Centre National de la Recherche Scientifique (CNRS), Commissariat \`a
l'\'ener\-gie atomique et aux \'energies alternatives (CEA),
Commission Europ\'eenne (FEDER fund and Marie Curie Program),
Institut Universitaire de France (IUF), IdEx program and UnivEarthS
Labex program at Sorbonne Paris Cit\'e (ANR-10-LABX-0023 and
ANR-11-IDEX-0005-02), Labex OCEVU (ANR-11-LABX-0060) and the
A*MIDEX project (ANR-11-IDEX-0001-02),
R\'egion \^Ile-de-France (DIM-ACAV), R\'egion
Alsace (contrat CPER), R\'egion Provence-Alpes-C\^ote d'Azur,
D\'e\-par\-tement du Var and Ville de La
Seyne-sur-Mer, France;
Bundesministerium f\"ur Bildung und Forschung
(BMBF), Germany; 
Istituto Nazionale di Fisica Nucleare (INFN), Italy;
Stichting voor Fundamenteel Onderzoek der Materie (FOM), Nederlandse
organisatie voor Wetenschappelijk Onderzoek (NWO), the Netherlands;
Council of the President of the Russian Federation for young
scientists and leading scientific schools supporting grants, Russia;
National Authority for Scientific Research (ANCS), Romania;
Mi\-nis\-te\-rio de Econom\'{\i}a y Competitividad (MINECO):
Plan Estatal de Investigaci\'{o}n (refs. FPA2015-65150-C3-1-P, -2-P and -3-P, (MINECO/FEDER)), Severo Ochoa Centre of Excellence and MultiDark Consolider (MINECO), and Prometeo and Grisol\'{i}a programs (Generalitat
Valenciana), Spain; 
Ministry of Higher Education, Scientific Research and Professional Training, Morocco.
We also acknowledge the technical support of Ifremer, AIM and Foselev Marine
for the sea operation and the CC-IN2P3 for the computing facilities.

\end{document}